\documentclass[12pt]{article}
\usepackage{epsfig}

\def\lsim{\raise0.3ex\hbox{$<$\kern-0.75em\raise-1.1ex\hbox{$\sim$}}}
\def\gsim{\raise0.3ex\hbox{$>$\kern-0.75em\raise-1.1ex\hbox{$\sim$}}}

\def\beq{\begin{equation}}
\def\eeq{\end{equation}}
\def\bea{\begin{eqnarray}}
\def\eea{\end{eqnarray}}
\def\noi{\noindent}
\begin{document}

%\rightline{\bf DRAFT}
\rightline{LPT Orsay 00-127}
\rightline{UCOFIS 2/00}
%\rightline{hep-ph/0011352}
%\rightline{\today}
\rightline{July 2001}

\vspace{1cm}
 
\begin{center}
{\Large\bf A geometrical estimation of saturation\\
of partonic densities}\vspace{1cm}
 
N. Armesto$^{a}$
and
C. A. Salgado$^{b}$
 
\vspace{0.2cm}
 
$^a$ {\it Departamento de F\'{\i}sica, M\'odulo C2, Planta baja, Campus
de Rabanales,}\\
{\it Universidad de C\'ordoba, E-14071 C\'ordoba, Spain} \\
 
$^b$ {\it Laboratoire de Physique Th\'eorique, Universit\'e de Paris
XI,}\\
{\it B$\hat{a}$timent 210, F-91405 Orsay Cedex, France}\\
 
\end{center}

\vspace{2cm}
{\small
We propose a new criterium for saturation of the density of partons both
in nucleons and nuclei. It is applicable to any multiple scattering
model
which would be used to compute the number of strings exchanged in $ep$
and
$eA$ collisions. The criterium is based on percolation of strings, and
the
onset of percolation is estimated from expectations coming from the
study
of heavy ion collisions at high energies. We interpret this onset as an
indication of saturation of the density of partons in the wave function
of
the hadron. In order to produce quantitative results, a particular model
fitted to describe present HERA data and generalized to the nuclear case
is used. Nevertheless, with the number of scatterings controlled by the
relation between inclusive and diffractive processes, conclusions are
weakly model-dependent as long as different models are tuned to describe
the experimental data. This constitutes a new approach, based on the
eikonal description of soft hadronic collisions, and different from
others
which employ either perturbative QCD ideas or semiclassical methods. It
offers an alternative picture for saturation in the small $Q^2$ region.
}

\vfill


\newpage

Much interest has recently been devoted to the saturation of partonic
densities \cite{glr}, i.e. the change in the increase of the partonic
densities from power-like to logarithmic or constant with decreasing
parton momentum fraction $x$, both in nucleons
\cite{wust,levin,model2} and in nuclei
\cite{mueller1,misha,semi}. From the point of view of
experimental data on lepton-hadron scattering,
the most striking feature was the change in the logarithmic slope
of the proton structure
function $dF_2/d\ln{(Q^2)}$ at $x\sim
10^{-4}$, the so-called Caldwell
plot \cite{caldwell} (now known to be mainly due to a $Q^2-x$ correlation),
but the situation is not conclusive: Nucleon data
can be described not only in approaches which
consider saturation \cite{wust,levin,model2}, but also
satisfactorily
accommodated in the usual global fits 
\cite{gprot} (also available
for nuclei \cite{gnucl}) 
that consider
the standard QCD evolution or resummation \cite{resum},
starting from initial conditions at low photon virtualities $Q^2$ which
do not include
saturation (see \cite{merino} for an application to the Caldwell plot
and \cite{recent} for a discussion on the present situation).

From a theoretical point of view, the saturation regime is a very
interesting one characterized by a small coupling constant and high
occupation numbers, where
a semiclassical description in terms of fields has been proposed
\cite{semi}. Different models
offer explanations
based on multiple scattering (i.e. unitarization) or gluon interaction, both
in the case of nucleons \cite{glr,wust,levin,model2} and nuclei
\cite{glr,mueller1,misha,semi}.
These two approximations to the problem are
equivalent (see e.g. \cite{mueller2}) in different reference frames,
but the models predict the
onset of saturation in different kinematical regions and the saturation
features are also diverse.
In this short note we will essay another approach to the problem,
inherited from multiparticle production in nucleus-nucleus ($AB$)
collisions
at high energies and applicable to any model formulated in terms of
multiple scatterings.

The concept of saturation, not of the density of partons in the
hadronic wave function but of the number of partons produced in the
collision, was proposed some time ago \cite{muebla} in
$AB$
collisions at high energies and has been reconsidered recently in the
context of the search of the Quark Gluon Plasma (QGP); such high partonic
density should provide the initial condition for the possible
thermalization of the created system. Several related ideas have been used to
compute the multiplicity of produced particles in $AB$ collisions at
the Relativistic Heavy
Ion Collider (RHIC) at BNL and at the future
Large Hadron Collider (LHC) at CERN
\cite{bass,review}. For example, in \cite{ekrt}
perturbative QCD (pQCD) is used to compute the initial number of gluons,
quarks and antiquarks, which are limited according to the simple
geometrical criterium that the number of partons per unit of transverse
space times their transverse
dimension ($\propto 1/p_\perp^2$) cannot be greater than 1 (see
\cite{pirner} for other attempts in this direction). Besides,
the semiclassical methods used in \cite{semi} have also been
employed to estimate the initial number of gluons in a heavy ion
collision \cite{kv}.

On the other hand and in the framework of string models for soft
multiparticle production (see \cite{dpm} and references therein), a
simple geometrical criterium for saturation has been proposed. In these models
particle production comes from string breaking, strings
which are considered, in a first
approximation, as formed and decayed independently. As the number of
binary nucleon-nucleon
collisions (each one producing 2 strings \cite{dpm,kaidalov})
increases with increasing centrality, energy or nuclear
mass,
this approximation should break down. The onset of this phenomenon can be
estimated considering strings with a certain area in the transverse
space of the collision, and taking into account the possibility of
two-dimensional percolation of the strings when they overlap in this
transverse space. Percolation is a second order phase transition which
takes place when clusters of overlapping strings, with a size of the
order of the total transverse area available, appear. This idea
has been proposed in $AB$ collisions
\cite{percol1} and applied to signatures of QGP \cite{percol2}.

The purpose of this note is to use percolation of strings as an
indication for the onset of saturation of the density of partons in nucleons
and nuclei,
a quantity which in our case is not directly related with the partonic
densities measured in DIS, as such identification
\cite{mueller1,mueller2} can only be done at
high $Q^2$, and our approach will be devoted to the low $Q^2$ regime.
For this, we need a multiple exchange model for $ep$
collisions which allows us to compute the number of binary collisions,
generalize it to the nuclear case, translate the number of collisions to
a number of strings and estimate the density of strings to compute
whether percolation takes place or not. The method can be applied to any
multiple scattering model, and the results in any of these models
should
be quite the same
(within the uncertainties due to the extrapolation of the model
to nuclei and to
higher energies or smaller $x$)
as long as the model is able to describe the
fully inclusive and diffractive
experimental data on $ep$ collisions, see comments below.

Let us give a brief description of the model developed in
\cite{model2}, which is the particular one we are going to use to compute
the number of binary collisions and then of strings in order to give
quantitative predictions.
The goal of the model was the description of total and diffractive 
data on $ep$ scattering at low and moderate $Q^2$ and small $x$. 
This region is where
unitarity corrections are more important and where the transition from
non-perturbative to perturbative QCD takes place. 

In the proton rest frame,
the virtual photon coming from the lepton fluctuates into a $q\bar q$ state.
Then this hadronic state interacts with the proton. Unitarity corrections are
described by the multiple scattering of the $q\bar q$ fluctuation with the
proton. In a quasieikonal approach the total cross section is given by
\beq
\label{eq1}
\sigma_{tot}^{\gamma^*p}(s,Q^2)
=4g(Q^2)\int d^2b \left ({1-\exp\{-C\chi(s,Q^2,b)\}\over 2C}\right ),
\eeq
where $g(Q^2)$ is the $\gamma^*-(q\bar q)$ coupling, 
$2\chi(s,Q^2,b)$ is the elementary $(q\bar q)-p$ cross section
at fixed impact parameter $b$,
and $C=1.5$
is a parameter taking into account the diffractive dissociation of the
proton.
For small sizes
$r$ of the $q\bar q$ pair, $\chi\propto r^2$ from pQCD calculations. As
$r^2\propto 
1/Q^2$, for these small sizes $\chi\propto 1/Q^2$. For large sizes of the
fluctuation no $Q^2$-dependence is expected. In
\cite{model2} two components, corresponding to small ($S$) and large
($L$) sizes of the $q\bar q$ pair, were
taken into account, $\chi=\chi_L+\chi_S$. 
The fact that $\chi_S\propto 1/Q^2$ while $\chi_L$ does not
depend on $Q^2$ makes the correction terms in Eq. (\ref{eq1}) more important
for the $L$ part than for the $S$ one. So, more scatterings are present
-- in 
average -- for the $L$ than for the $S$ component;
for this reason and
also due to the fact that we will consider small $Q^2$, only the
$L$ component,
which
is the dominant one \cite{model2}
for $Q^2\ \lsim\  2$ (GeV/c)$^2$,
will be used in the actual computations.
The energy dependence of 
these $\chi$'s is given by a single pomeron of intercept
$\Delta=0.2$,
\beq
\label{eq2}
\chi_L={C_L\over \lambda_L} 
\exp\left\{ \Delta \xi -{b^2\over 4\lambda_L}\right\}.
\eeq
Here, $\xi=\ln {s+Q^2\over s_0+Q^2}$,
$\lambda_L=R^2_L+\alpha_P^\prime\xi$ with $R^2_L=3$
GeV$^{-2}$, and 
$\alpha_P^\prime=0.25$ GeV$^{-2}$, the slope of the pomeron trajectory,
gives the $\ln s$ behavior of the total cross section for very large
$s$; besides, $C_L=0.56$ GeV$^{-2}$ and $s_0=0.79$ GeV$^{2}$.
The variable $\xi$ is chosen so that $\chi\propto
x^{-\Delta}$ for large $Q^2\gg s_0$ and
$\chi\propto (s/s_0)^{\Delta}$
for $Q^2\to 0$; in this way, the model can be used
for photoproduction.
For the $S$ part, similar expressions were used in \cite{model2}
with an extra $r^2$ factor in Eq. (\ref{eq2}).

The description of diffraction is a very important ingredient of the model.
It is given by quadratic and higher order terms in $\chi$ in the expansion
of Eq. (\ref{eq1}). Thus, 
the ratio $\sigma_{diff}/\sigma_{tot}$ controls the 
unitarity (multiple scattering) corrections, i.e. the number of
scatterings and strings; this idea has been used to compute
nuclear structure functions from a description of diffraction at HERA,
see e.g. \cite{funnuc}.
So, any multiple scattering model able
to reproduce
the experimental data on this ratio should produce roughly the same
number of scatterings (strings) and, consequently, the same predictions
for the onset of percolation and saturation. A triple pomeron term was
introduced in \cite{model2} in order to reproduce large mass
diffractive processes. This term is another source of shadowing corrections
to the total cross section.
It will be used in the actual computations,
see \cite{model2} for the full expressions and parameters. The reggeon
contribution which appears in \cite{model2}
decreases with increasing energy and is negligible at the
energies under consideration, so it has been ignored.

The model in \cite{model2} has 9 free parameters that were fitted
to experimental data on diffractive and total $ep$ cross sections for
$0\le Q^2\ \lsim\  10$ (GeV/c)$^2$ and $10^{-6}\ \lsim\  x\ \lsim\  10^{-2}$.
Once the parameters of the model are fitted, it is possible to know the 
mean number of collisions \cite{kaidalov}:
\bea
\label{eq3}
\bar n&=&\sum_{n=1}^\infty \ n\ {\int d^2b
\ \sigma_n(s,Q^2,b)\over\sum_{n=1}^\infty
\int d^2b^\prime\ \sigma_n(s,Q^2,b^\prime)}\nonumber \\
&=&\frac{\int d^2b \
2C\chi(s,Q^2,b)}{\int d^2b^\prime
[1-\exp{\{-2C\chi(s,Q^2,b^\prime)\}}]}\ ,
\eea
where, for $n\ge 1$,
\beq
\label{eq4}
\sigma_n(s,Q^2,b)={g(Q^2)\over C}{[2C\chi(s,Q^2,b)]^n\over n!}
\exp(-2C\chi).
\eeq
Notice that in these expressions, $\chi(s,Q^2,b)$ contains the triple pomeron 
contribution, so cuts in different branchings of
one single fan diagram (i.e. one single
tree of triple pomeron couplings)
are included in the same $\sigma_n$, which thus corresponds to the exchange of
$n$ fan diagrams, each of them
cut in one o more than one of its branches.
Thus Eq. (\ref{eq4}) is
a conservative estimation, as
these cuts
could give rise to a larger
number of strings. Besides, all our expressions are asymptotic
ones, not considering energy-momentum conservation (which could
reduce slightly the number of collisions at the lowest energies).

Neglecting isospin at high energies,
the generalization of any multiple scattering model formulated for $ep$
to the case of $eA$ collisions is straightforward in the
Glauber-Gribov approach \cite{glaugri}:
The number of $q\bar q- {\rm nucleon}$
collisions (the number of participating nucleons
of $A$) in this case, is given \cite{glau}
in terms of the inelastic non-diffractive
cross sections by
\beq
\label{eq7}
\langle n_{part}\rangle=A\ \sigma_{in}^{\gamma^*p}/
\sigma_{in}^{\gamma^*A}\propto A^{1/3}
,
\eeq
with
\beq
\label{eq8}
\sigma_{in}^{\gamma^*A}=g(Q^2)\int d^2b \left(1-\exp{\left\{-A
T_A(b)\sigma_{in}^{\gamma^*p}/g(Q^2)
\right\}}\right),
\eeq
$T_A(b)=\int_{-\infty}^\infty
dz \rho_A(z,\vec{b})$ the profile function normalized
to 1 taken from
\cite{prof} and
\bea
\label{eq9}
\sigma_{in}^{\gamma^*p}&=& \sum_{n=1}^\infty
\int d^2\tilde b \ \sigma_n(s,Q^2,\tilde b)\nonumber \\
&=&{g(Q^2)\over C}
\int d^2\tilde b \ [1-\exp{\{-2C\chi(s,Q^2,\tilde b)\}}].
\eea
So, the total number of collisions is given by $\langle
n_{coll}\rangle=\langle n_{part}\rangle\ \bar n$. As previously
commented, in the actual
computations we will only use the $L$ component, Eq. (\ref{eq2}), as it
is the dominant one \cite{model2}
for $Q^2\ \lsim\  2$ (GeV/c)$^2$ where our calculations
will be done.

At this point, it could be argued that using the model in \cite{model2}
(or any other multiple scattering model)
there is a possibility to study saturation of the density of partons,
both in nucleons and in nuclei,
simply looking at the point in which amplitudes in impact parameter
space become energy independent, or alternatively
the point in which cross sections reach a regime
in which their energy behavior becomes identical to that of the size of
the target (expanding in the case of a nucleon).
Nevertheless, the generalization of
\cite{model2} to nuclei is not so obvious: ours is a very simple
one,
but more rigorous generalizations \cite{nucleus}
also rely in simplifications of the exact Gribov calculus \cite{regge}
or Glauber-Gribov theory \cite{glaugri}. So we think that
an estimation as the one we perform, based on geometrical criteria, is
worthy, of simple and general applicability,
and may provide, as in the case of nucleus-nucleus collisions,
an
indication of the onset of a high density, non-linear regime.

Let us establish now our criterium for saturation
of the density of
partons in the wave function of the target.
As it was said,
percolation is a non-thermal second order phase transition, which takes
place when clusters of overlapping objects acquire a size comparable to
the total size available \cite{isichenko}.
In our case the space is the transverse
dimension available for the collision, and the overlapping objects are
strings. The parameter which controls the onset of percolation is the
dimensionless string density
\beq
\eta=N\ t/T
\label{eq10}
\eeq
(which may be related \cite{review} with the dimensionless density of
gluons found
in semiclassical models),
with
\beq
N=2\ \langle
n_{coll}\rangle
\label{eq11}
\eeq
the number of strings exchanged in the collision (each
collision gives rise to two strings due to the pomeron dominance at
high energies, see \cite{dpm,kaidalov}), $t=\pi r_0^2$ is the transverse
dimension of the string\footnote{In this approach
this transverse dimension plays the
r$\hat{\mbox o}$le of an intrinsic scale of soft physics, $Q^2 \to 0$,
where a description in terms of pQCD degrees of freedom becomes
dubious.},
with $r_0\simeq 0.20\div 0.25$ fm as extracted
from phenomenology \cite{percol1,percol2}, and $T$ the total
transverse area available for the collision. 
This last quantity is not known and could depend on the virtuality of the
fluctuation $Q^2$; however, for small and moderate $Q^2$, it can
be estimated to be the typical size of the vector meson in which the 
virtual photon fluctuates (as this is the smaller object in the interaction). 
So, we will use $T=1$ fm$^2$ (a radius $\sqrt{T/\pi}\simeq 0.56$ fm). 
Also, a size varying with the energy, in the spirit of an expanding
proton, could be explored; for example, a size increasing with increasing
energy would slow the corresponding increase of the density of strings but,
for simplification, we will
employ a fixed size.

The critical value for $\eta$ where percolation takes place,
has been computed using different methods
and depends quite strongly on the profile of the nucleus (i.e. on the
distribution of the overlapping objects inside the available transverse
space). For continuum
two-dimensional percolation and from
\cite{isichenko,dias}, we take $\eta_c\simeq 1.12\div 1.50$. Defining the string
density as $n=N/T$
and allowing
for the different values of $r_0$ and $\eta_c$,
we find a critical string density
\beq
n_c\simeq 6\div 12 \ \ {\rm strings/fm}^2.
\label{eq12}
\eeq

With this critical string density,
it is tempting
to estimate the behavior of the $Q^2$ at which, for a fixed $s$,
saturation of the density of partons
takes place in this approach, $Q^2_{sat}$
\cite{glr,wust,levin,mueller1,misha,semi,mueller2}.
However, in our case, being $\chi_L$ almost independent on $Q^2$, the density
of strings is also almost $Q^2$-independent for fixed $s$; besides, the
model is only valid for small $Q^2$ and has not been designed for
$Q^2$-evolution.
More significant in our model is the value of $x$ or $s$ where, for
fixed $Q^2$ and $A$, saturation takes place, $x_{sat}$ or $s_{sat}$
respectively.
Considering
neither triple pomeron nor reggeon contributions and approximating in Eq.
(\ref{eq7}) $\sigma_{in}^{\gamma^*A}\simeq
g(Q^2)\ \pi R_0^2 A^{2/3}$, with $R_0\simeq 1.2$ fm, 
it is found that $x_{sat}
\propto [Q^2/(s_0+Q^2)]\ A^{1/(3\Delta)}$
and $s_{sat}\propto (s_0+Q^2)
\ A^{-1/(3\Delta)}$, $1/(3\Delta)=5/3$.
So, for $Q^2\to 0$, $x_{sat}$ increases linearly with increasing $Q^2$
while $s_{sat}$ is roughly constant; these qualitative features will be
observed in the numerical
results.

Let us turn to numerical evaluations.
Using Eqs. (\ref{eq3})-(\ref{eq9}),
(\ref{eq11}) and (\ref{eq12}), we can now
compute the string density for different hadronic targets, $Q^2$, and
$x$ or $s$. When this density becomes larger than the critical value, Eq.
(\ref{eq12}), percolation will take place, which we will interpret as a
signal of the onset of saturation of the density of partons in the target.
In Fig. 1 we present results for
the string density in $\gamma^* - A$ collisions versus $x$
for
$Q^2=0.1$ and 1  (GeV/c)$^2$, with $A=1$, 9, 56 and 207
(corresponding to $p$, $Be$, $Fe$ and $Pb$ respectively).
In Fig. 2 the same quantity is presented
versus $s$
for
$Q^2=0$ and 2 (GeV/c)$^2$.
Some comments are in order: First, from
Fig. 1 it looks as if saturation (percolation)
is favored by a higher $Q^2$,
apparently in contradiction to what is commonly expected. This is due to
the fact that, as we have said, the number of rescatterings is hardly
dependent on $Q^2$ -- as can also
be observed in Fig. 2, and that $s$ is the
variable which controls this number
(indeed, in Fig. 2 it can be
seen that $s_{sat}$ increases slightly
with increasing $Q^2$, as expected).
So, for a fixed $s$
at which percolation occurs, the higher the $Q^2$ the higher the
$x_{sat}$
(in agreement with the naive
expectations in the previous paragraph). Second, the dependence of
$s_{sat}$ on $A$, parametrized as $A^{-\alpha}$, is found to be stronger in
the numerical computations ($\alpha\simeq 5/2$) than the power 5/3
estimated in the previous paragraph. This discrepancy is due
to the triple pomeron contribution included in the numerical
computations, which appear as a denominator in Eq. (\ref{eq2}),
diminishing the 'effective' $\Delta$ which appears in $s_{sat}$ and thus
making $\alpha=1/(3\Delta)$ larger.

To summarize, a criterium for saturation in the small $Q^2$ region
applicable to any multiple scattering model, has been presented.
To produce quantitative results, a
multiple scattering model for $\gamma^* -p$ collisions
in this $Q^2$ region \cite{model2}
has been generalized to the nuclear case, and used to compute the number of
exchanged strings. 
As multiple scattering (and thus the number of produced strings) is
controlled by the ratio $\sigma^{diff}/\sigma^{tot}$, this number is
related to experimental data on diffraction and the actual realization 
of the model is not crucial to compute the string densities as long as
it reproduces the experimental data.
Employing the ideas
of percolation of strings taken from Heavy Ion Physics, the kinematical
regions for the onset of
percolation, which has been interpreted as
saturation of the density of partons in the target\footnote{As
commented previously,
we are working in
the low $Q^2$ regime, so this quantity cannot be identified with the partonic
densities measured in DIS.},
have been calculated.
This constitutes a new approach, based
on Regge phenomenology and different from others which use
either pQCD ideas or semiclassical methods; it offers an alternative
picture, based in hadronic degrees of freedom (strings),
for saturation in the small $Q^2$ region. In view of the results
presented in the Figures for the onset of percolation, which for large
nuclei may appear at not so small $x$, saturation could be
observed in future $eA$ colliders \cite{future}, and the effects of this
second order phase
transition visible in correlations (as
proposed
in $AB$ collisions \cite{fluctu}).

\vskip 1cm

\noi {\bf Acknowledgments:}
The authors express their gratitude to A. Capella
for a critical reading of the manuscript, to K. J. Eskola for useful
discussions on the model of saturation in \cite{ekrt}, and
to C. Pajares for his interest in
this work and constant encouragement.
N. A. acknowledge financial
support by CICYT of Spain under contract
AEN99-0589-C02 and by Universidad de C\'ordoba,
and C. A. S. a postdoctoral grant
by Ministerio de Educaci\'on y Cultura of Spain.
Laboratoire de Physique Th\'eorique is Unit\'e Mixte de Recherche -- CNRS
-- UMR n$^{\rm o}$ 8627.

\newpage
\centerline{\bf \Large Figure captions:}
\vskip 1cm
 
\noi {\bf Fig. 1.} 
String density in $\gamma^* - p$, $Be$, $Fe$ and $Pb$ collisions versus
$x$ for $Q^2=0.1$ (GeV/c)$^2$ (solid lines) and $Q^2=1$ (GeV/c)$^2$ (dashed 
lines). Dotted lines are the bounds on the critical string density for 
percolation of strings.

\vskip 0.5cm
\noi {\bf Fig. 2.} 
String density in $\gamma^* - p$, $Be$, $Fe$ and $Pb$ collisions versus
$s$ for photoproduction (solid lines) and $Q^2=2$ (GeV/c)$^2$ (dashed 
lines). Dotted lines are the bounds on the critical string density for 
percolation of strings.

\newpage
\centerline{\bf \Large Figures:}
 
\vskip 3cm
 
\begin{center}
\epsfig{file=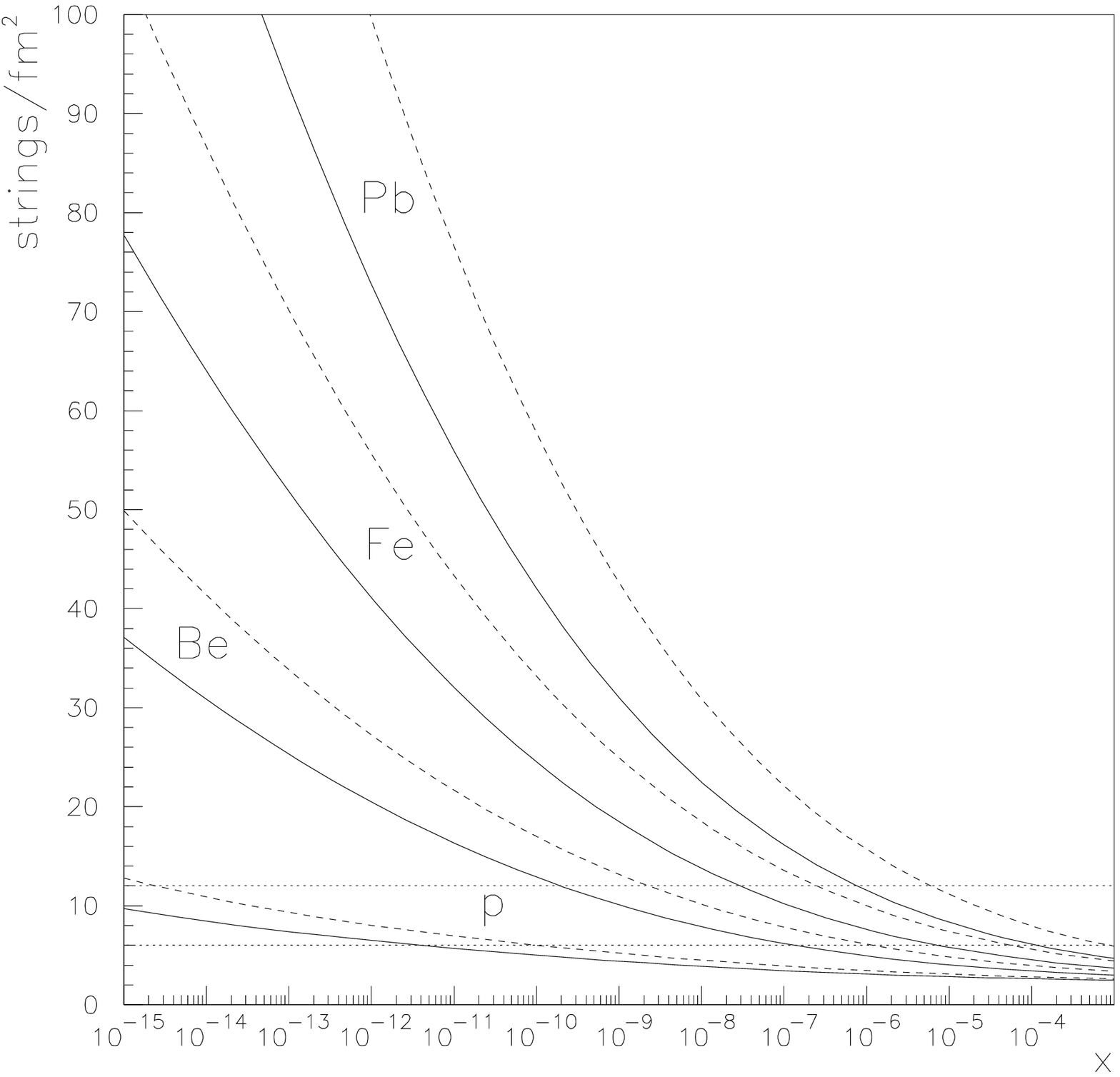,width=15.5cm}
\vskip 1cm
{\bf \large Fig. 1}
\end{center}
 
\newpage
 
\begin{center}
\epsfig{file=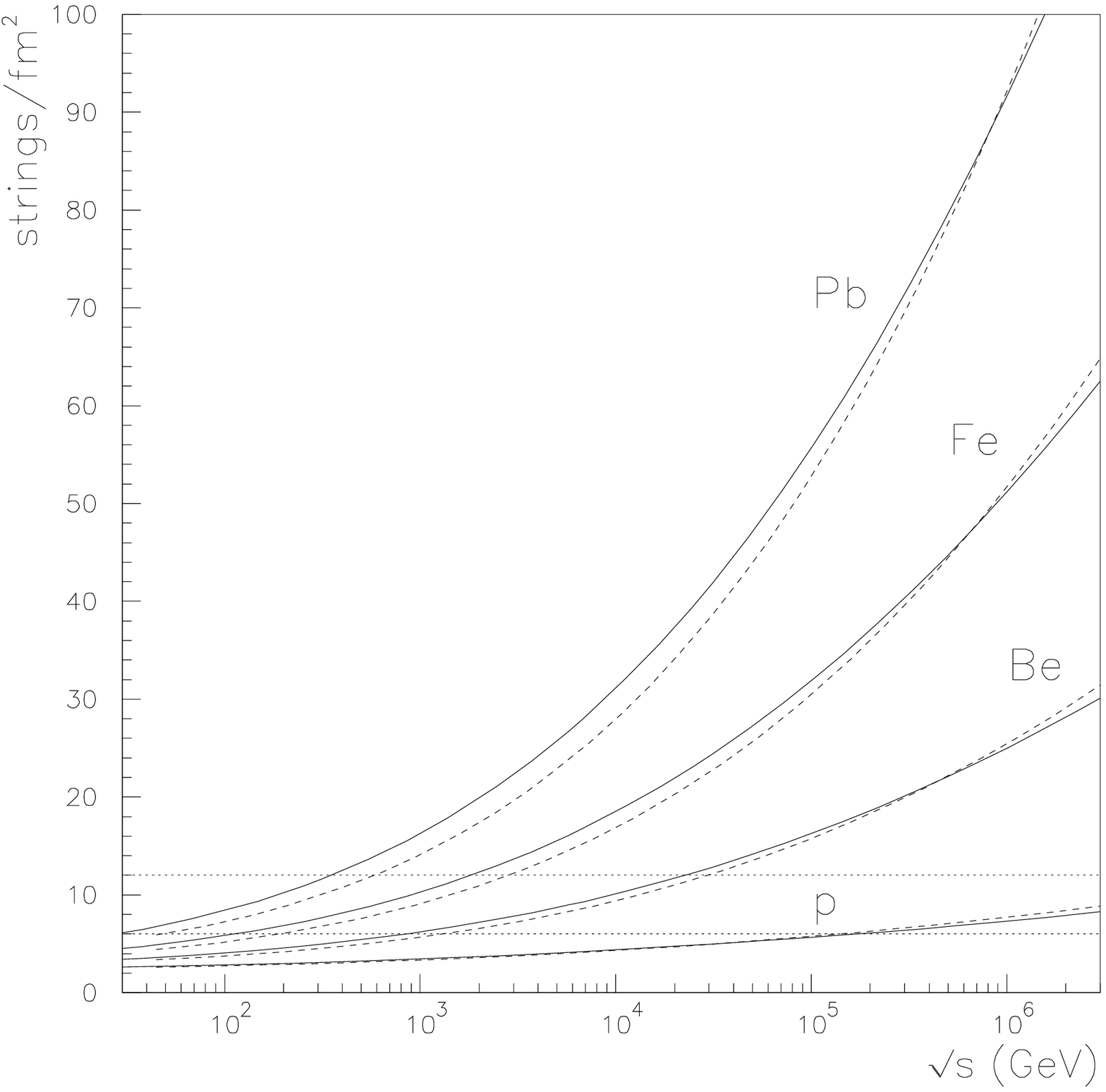,width=15.5cm}
\vskip 1cm
{\bf \large Fig. 2}
\end{center}


\begin{thebibliography}{99}

\bibitem{glr} L. V. Gribov, E. M. Levin and M. G. Ryskin,
Phys. Rept. {\bf 100}, 1 (1983);
A. H. Mueller and J.-W. Qiu, Nucl. Phys. {\bf B268}, 427 (1986). 

\bibitem{wust} K. Golec-Biernat and M. W\"usthoff,
Phys. Rev. {\bf D59}, 014017 (1999); {\bf D60}, 114023 (1999).

\bibitem{levin}
E. Gotsman, E. M. Levin,
U. Maor and E. Naftali, Nucl. Phys. {\bf B539}, 535 (1999);
E. M. Levin and U. Maor, preprint TAUP-2643-2000 (hep-ph/0009217);
M. B. Gay Ducati and V. P. Gon\c calves,
Phys. Lett. {\bf B466}, 375 (1999).

\bibitem{model2} A. Capella, E. G. Ferreiro, A. B. Kaidalov and C. A.
Salgado,
Nucl. Phys. {\bf B593}, 336 (2001);
Phys. Rev. {\bf D63}, 054010 (2001).

\bibitem{mueller1} A. H. Mueller, Nucl. Phys. {\bf B335}, 115 (1990);
{\bf B558}, 285 (1999);
Yu. V. Kovchegov and A. H. Mueller,
Nucl. Phys. {\bf B529}, 451 (1998); Yu. V. Kovchegov, Phys. Rev. {\bf
D54}, 5463 (1996); {\bf D55}, 5445 (1997);
E. M. Levin and K. Tuchin,
Nucl. Phys. {\bf B573}, 383 (2000); E. M. Levin and M. Lublinsky,
preprint TAUP-2670-2001 (hep-ph/0104108).

\bibitem{misha} M. A. Braun,
Eur. Phys. J. {\bf C16}, 337 (2000); hep-ph/0010041; N. Armesto
and M. A. Braun, Eur. Phys. J. {\bf C20}, 517 (2001).

\bibitem{semi}
L. McLerran and R. Venugopalan, Phys. Rev. {\bf D49}, 2233 (1994); 3352;
{\bf D50}, 2225 (1994);
J. Jalilian-Marian, A. Kovner, L. McLerran and
H. Weigert, Phys. Rev. {\bf D55}, 5414 (1997);
E. Iancu, A. Leonidov and L. McLerran, preprint
Saclay-T00/166 and BNL-NT-00/24
(hep-ph/0011241); E. Iancu and L. McLerran, Phys. Lett. {\bf B510}, 145
(2001).

\bibitem{caldwell} A. Caldwell at the {\it DESY Theory Workshop}
(Hamburg,
Germany,
October 1997); ZEUS Collaboration: J. Breitweg {\it et al.},
Eur. Phys. J. {\bf C7}, 609 (1999).

\bibitem{gprot} A. D. Martin, R. G. Roberts,
W. J. Stirling and R. S. Thorne,
Eur. Phys. J. {\bf C4}, 463 (1998);
M. Gl\"uck, E. Reya and A.
Vogt,
Eur. Phys. J. {\bf C5}, 461 (1998);
L. Lai {\it et al.},
Eur. Phys. J. {\bf C12}, 375 (2000).

\bibitem{gnucl} K. J. Eskola, V. J. Kolhinen and C. A. Salgado,
Eur. Phys. J. {\bf C9}, 61 (1999);
D. Indumathi and W. Zhu,
Z. Phys. {\bf C74}, 119 (1997); M. Hirai, S. Kumano and M.
Miyama, Phys. Rev. {\bf D64}, 034003 (2001).

\bibitem{resum} G. Altarelli, R. D. Ball and S. Forte,
Nucl. Phys. {\bf B599}, 383 (2001);
R. S. Thorne,
Nucl. Phys. {\bf B512}, 323 (1998); M. Ciafaloni, D. Colferai and
G. P. Salam,
Phys. Rev. {\bf D60}, 114036 (1999).

\bibitem{merino} A. B. Kaidalov, C. Merino and D. Pertermann,
Eur. Phys. J. {\bf C20}, 301 (2001).

\bibitem{recent} E. Gotsman, E. Ferreira, E. M. Levin,
U. Maor and E. Naftali, Phys. Lett. {\bf B500}, 87 (2001).

\bibitem{mueller2} A. H. Mueller, in {\it Proceedings of the XVII Autumn
School: QCD: Perturbative or Nonperturbative?}, Eds. L. S. Ferreira, P.
Nogueira and J. I. Silva-Marcos, World Scientific, Singapore 2001, p.
180.

\bibitem{muebla} J. P. Blaizot and A. H. Mueller,
Nucl. Phys. {\bf B289}, 847 (1987); A. H. Mueller,
Nucl. Phys. {\bf
B572}, 227 (2000).

\bibitem{bass} S. A. Bass {\it et al.}, Nucl. Phys. {\bf A661}, 205c (1999).

\bibitem{review} N. Armesto and C. Pajares, Int. J. Mod. Phys. {\bf
A15}, 2019 (2000).

\bibitem{ekrt}
K. J. Eskola, K. Kajantie, P. V. Ruuskanen and K. Tuominen,
Nucl. Phys. {\bf B570}, 379 (2000);  K. J. Eskola, K. Kajantie and K. Tuominen,
Phys. Lett. {\bf B497}, 39 (2001); K. J. Eskola, P. V. Ruuskanen, S. S.
R\"as\"anen and K. Tuominen, preprint JYFL-3/01 (hep-ph/0104010).

\bibitem{pirner} D. Kharzeev and M. Nardi,
Phys. Lett. {\bf B507}, 121 (2001);
H.-J. Pirner and F. Yuan, Phys. Lett. {\bf B512}, 297 (2001).

\bibitem{kv} A. Krasnitz and R. Venugopalan,
Phys. Rev. Lett. {\bf 84}, 4309 (2000); {\bf 86}, 1717 (2001).

\bibitem{dpm} A. Capella, U.-P. Sukhatme,
C.-I. Tan and J. Tran Thanh Van,
Phys. Rept. {\bf 236}, 225 (1994).

\bibitem{kaidalov} A. B. Kaidalov, Sov. J. Nucl. Phys. {\bf 45}, 902
(1987); Yu. M. Shabelsky, Z. Phys. {\bf C57}, 409 (1993).

\bibitem{percol1} N. Armesto, M. A. Braun, E. G. Ferreiro and C.
Pajares,
Phys. Rev. Lett. {\bf 77}, 3736 (1996).

\bibitem{percol2} M. Nardi and H. Satz, Phys. Lett. {\bf B442}, 14 (1998); H.
Satz, Nucl. Phys. {\bf A642}, 130 (1998);
J. Dias de Deus, R. Ugoccioni and A. Rodrigues,
Eur. Phys. J. {\bf C16}, 537 (2000).

\bibitem{funnuc} A. Capella, A. B. Kaidalov, C. Merino, D. Pertermann
and J.
Tran Thanh Van,
Eur. Phys. J. {\bf C5}, 111 (1998).

\bibitem{glaugri} R. J. Glauber, in {\it Lectures in Theoretical
Physics}, Vol. 1, ed. W. E. Brittin and L. G. Duham
(Interscience, New York, 1959);
V. N. Gribov, Sov. Phys. JETP {\bf 29}, 483 (1969);
{\bf 30}, 709 (1970).

\bibitem{glau} A. Bialas, M. Bleszy\'nski and
W. Czyz,
Nucl. Phys. {\bf B111}, 461 (1976).

\bibitem{prof} C. W. De Jager, H. De Vries and C. De Vries,
Atom. Data Nucl. Data Tabl. {\bf 14}, 479 (1974).

\bibitem{nucleus}
A. Schwimmer,
Nucl. Phys. {\bf B94}, 445 (1975);
L. Caneschi, A. Schwimmer and R.
Jengo,
Nucl. Phys. {\bf B108}, 82 (1976);
A. Capella, A. B. Kaidalov and
J. Tran Thanh Van,
Heavy Ion Phys. {\bf 9}, 169 (1999); S. Bondarenko, E. Gotsman, E. M. Levin and
U. Maor,
Nucl. Phys. {\bf A683}, 649 (2001).

\bibitem{regge} V. N. Gribov, Sov. Phys. JETP {\bf 26}, 414 (1968).

\bibitem{isichenko} M. B. Isichenko, Rev. Mod. Phys. {\bf 64}, 961
(1992).

\bibitem{dias} A. Rodrigues, R. Ugoccioni and J. Dias de Deus,
Phys. Lett. {\bf B458}, 402 (1999).

\bibitem{future} M. Arneodo {\it et al.},
in {\it Proceedings of the Workshop on Future Physics at
HERA} (Hamburg,
Germany,
September 1995); H. Abramowicz {\it et al.}, {\it
TESLA Technical Design Report, Part VI, Chapter 2}, Eds. R. Klanner, U.
Katz, M. Klein and A. Levy.

\bibitem{fluctu} M. A. Braun and C. Pajares,
Phys. Rev. Lett. {\bf 85}, 4864 (2000);
Yu. V. Kovchegov, E. M. Levin and
L. McLerran,
Phys. Rev. {\bf C63}, 024903 (2001).

\end{thebibliography}
\end{document}